\documentclass[runningheads]{llncs}

\newcommand{\tool}[1]{\textsc{#1}\xspace}
\newcommand{\toolset}{\tool{Modest Toolset}}
\newcommand{\mcsta}{\textsf{mcsta}\xspace}
\newcommand{\sepref}{\textsf{sepref}\xspace}
\newcommand{\isabelle}{\textsf{Isabelle/HOL}\xspace}
\newcommand{\isa}{\textsf{Isabelle}\xspace}

\newcommand{\Csharp}{\settoheight{\dimen0}{C}C\kern-.05em \resizebox{!}{\dimen0}{\raisebox{\depth}{\#}}\xspace}

\renewcommand{\emptyset}{\ensuremath{\varnothing}}

\newcommand*{\eg}{e.g.\@\xspace}
\newcommand*{\ie}{i.e.\@\xspace}

\newcommand{\tuple}[1]{\ensuremath{\langle #1 \rangle}}

\newcommand{\defeq}{\mathrel{\vbox{\offinterlineskip\ialign{\hfil##\hfil\cr{\tiny \rm def}\cr\noalign{\kern0.30ex}$=$\cr}}}}

\newcommand{\sunderscore}{\text{\scriptsize{\mathunderscore}}}

\newcommand{\mdp}{M} 
\newcommand{\states}{S} 
\newcommand{\kernel}{K} 
\newcommand{\pmf}{\mathcal{P}} 
\newcommand{\pol}{\pi} 

\newcommand{\target}{U} 
\newcommand{\prfin}[3]{\mathbb{P}_{#2}^{\le #3}(\lozenge #1)} 
\newcommand{\safetyfin}[3]{\mathbb{P}_{#2}^{\le #3}(\square #1)} 

\newcommand{\levels}{I}

\newcommand{\reach}{s_+}
\newcommand{\avoid}{s_-}

\newcommand{\nan}{NaN\xspace}
\newcommand{\avx}{AVX512\xspace}
\newcommand{\lbf}{\ensuremath{\mathit{lb}}\xspace}
\newcommand{\ubf}{\ensuremath{\mathit{ub}}\xspace}

\newcommand{\refcomp}[1][R]{\ensuremath{\leq_{#1}}\xspace}

\newcommand{\Rlb}{\ensuremath{R_{\mathit{lb}}}}

\newcommand{\Rid}{\ensuremath{R_{\mathit{id}}}\xspace}
\newcommand{\Alb}{\ensuremath{A_{\mathit{lb}}}}
\newcommand{\Aub}{\ensuremath{A_{\mathit{ub}}}}

\newcommand{\Asize}{\ensuremath{A_{\mathit{size}}}\xspace}

\usepackage[T1]{fontenc}
\usepackage[utf8x]{inputenx}

\usepackage{pgfplots}
\pgfplotsset{compat=1.14}
\usetikzlibrary{calc,angles,quotes,automata,arrows,decorations.pathmorphing,patterns,positioning,fit,trees,shapes,shapes.misc,shadows}

\usepackage{aligned-overset}
\usepackage{amsmath}
\usepackage{amssymb}
\usepackage{stmaryrd}
\usepackage{color}
\usepackage[inline]{enumitem}
\usepackage{float}
\usepackage{wrapfig}
\usepackage{lineno}
\usepackage{lstautogobble}
\usepackage{tikz}
\usepackage[nospace]{cite}
\usepackage{xifthen}
\usepackage{xspace}
\usepackage{etoolbox}
\usepackage{colonequals}
\usepackage{graphicx} 
\usepackage{hyperref} 
\usepackage[firstpage]{draftwatermark} 

\setitemize{noitemsep,topsep=0pt,parsep=0pt,partopsep=0pt,leftmargin=12.0pt}

\usepackage{todonotes}

\usepackage{cleveref}

\modulolinenumbers[5]

\makeatletter
\lst@AddToHook{OnEmptyLine}{\vspace{-0.8\baselineskip}}
\makeatother

\Crefname{figure}{Fig.}{Figs.}
\crefname{figure}{fig.}{figs.}
\Crefname{tabular}{Tab.}{Tabs.}
\crefname{tabular}{tab.}{tabs.}
\Crefname{section}{Sect.}{Sects.}
\crefname{section}{sect.}{sects.}
\Crefname{equation}{Eq.}{Eqs.}
\crefname{equation}{eq.}{eqs.}
\Crefname{definition}{Def.}{Defs.}
\crefname{definition}{def.}{defs.}
\Crefname{theorem}{Thm.}{Thms.}
\crefname{theorem}{thm.}{thms.}

\definecolor{color1}{RGB}{55,126,184}
\definecolor{color2}{RGB}{228,26,28}
\definecolor{color3}{RGB}{77,175,74}
\definecolor{color4}{RGB}{172,183,45}
\definecolor{color5}{RGB}{255,127,0}

\newboolean{anonymize}
\setboolean{anonymize}{false}
\newboolean{hideplots}
\newlength{\scatterplotsize}
\lstdefinelanguage{isabelle}{
  morekeywords={theorem,theorems,corollary,lemma,lemmas,locale,sublocale,global_interpretation,begin,end,fixes,assumes,shows,and,class,
    constrains , definition, defines, where, apply, done,unfolding, primrec, fun, using, by, for, uses, file,
    schematic_lemma, concrete_definition, prepare_code_thms, export_code, datatype, type_synonym, typedef, value,
    proof, next, qed, show, have, hence, thus, interpretation, fix, context, sepref_definition,is,export_llvm
 } ,
  morekeywords=[2]{rec, return, bind, foreach, if, then, else, do, let, in, spec, fail, assert, assume, while, case, of,
    check,with_split,npar,nseq},
  sensitive=True,
  morecomment=[s]{(\*}{\*)},
}

\DeclareTextCommand{\shortunderscore}{T1}{%
  \leavevmode \kern.06em\vbox{\hrule width.4em}}

\newcommand{\is}{\lstinline[language=isabelle]}

\counterwithin{isalemma}{section}
\makeatletter
\patchcmd{\lst@MakeCaption}{{lstlisting}}{{\verborgh@counter}}{}{}%
\def\verborgh@counter{lstlisting}
\def\verborgh@prefix{Lemma}
\AtBeginDocument{%
  \patchcmd{\theHlstnumber}{\thelstnumber}{\verborgh@prefix\thelstnumber}{}{}%
}

\lstnewenvironment{isalemma}[1][]
 {%
  \def\verborgh@counter{isalemma}%
  \def\verborgh@prefix{L}%
  \lstset{#1}%
 }
 {}
\makeatother

\crefname{isalemma}{lemma}{lemmas}
\Crefname{isalemma}{Lemma}{Lemmas}

\counterwithin{isatheorem}{section} 
\makeatletter
\patchcmd{\lst@MakeCaption}{{lstlisting}}{{\verborgh@counter}}{}{}%
\def\verborgh@counter{lstlisting}
\def\verborgh@prefix{D}
\AtBeginDocument{%
  \patchcmd{\theHlstnumber}{\thelstnumber}{\verborgh@prefix\thelstnumber}{}{}%
}

\lstnewenvironment{isatheorem}[1][]
 {%
  \def\verborgh@counter{isatheorem}%
  \def\verborgh@prefix{T}%
  \lstset{#1}%
 }
 {}
\makeatother

\crefname{isatheorem}{thm.}{thms.}
\Crefname{isatheorem}{Thm.}{Thms.}

\counterwithin{isadefinition}{section} 
\makeatletter
\patchcmd{\lst@MakeCaption}{{lstlisting}}{{\verborgh@counter}}{}{}%
\def\verborgh@counter{lstlisting}
\def\verborgh@prefix{D}
\AtBeginDocument{%
  \patchcmd{\theHlstnumber}{\thelstnumber}{\verborgh@prefix\thelstnumber}{}{}%
}

\lstnewenvironment{isadefinition}[1][]
 {%
  \def\verborgh@counter{isadefinition}%
  \def\verborgh@prefix{L}%
  \refstepcounter{theorem} 
  \lstset{#1}%
 }
 {}
\makeatother

\crefname{isadefinition}{def.}{defs.}
\Crefname{isadefinition}{Def.}{Defs.}

\counterwithin{isalocale}{section} 
\makeatletter
\patchcmd{\lst@MakeCaption}{{lstlisting}}{{\verborgh@counter}}{}{}%
\def\verborgh@counter{lstlisting}
\def\verborgh@prefix{L}
\AtBeginDocument{%
  \patchcmd{\theHlstnumber}{\thelstnumber}{\verborgh@prefix\thelstnumber}{}{}%
}

\lstnewenvironment{isalocale}[1][]
 {%
  \def\verborgh@counter{isalocale}%
  \def\verborgh@prefix{L}%
  \lstset{#1}%
 }
 {}
\makeatother

\crefname{isalocale}{locale}{locales}
\Crefname{isalocale}{Locale}{Locales}

\SetWatermarkAngle{0}
\SetWatermarkText{\raisebox{15cm}{
\hspace{-0.25cm}
\href{https://doi.org/10.4121/bf0fef24-4f0f-4de6-a58d-07b9ba601804}{\includegraphics{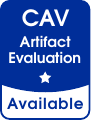}}
\hspace{8.3cm}
\includegraphics{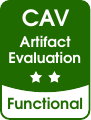}
}}

\newboolean{includeAppendix}
\setboolean{includeAppendix}{false}

\makeatletter
\def\orcidID#1{\smash{\href{http://orcid.org/#1}{\protect\raisebox{-1.25pt}{\protect\includegraphics{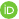}}}}}
\makeatother

\begin{document}
\title{A Formally Verified IEEE 754 Floating-Point Implementation of Interval Iteration for MDPs}
\titlerunning{Formally Verified Floating-Point Implementation of Interval Iteration}
%

\author{Bram Kohlen\inst{1}\orcidID{0000-0003-2908-8838} \and
Maximilian Sch{\"a}ffeler\inst{2}\orcidID{0000-0002-2612-2335} \and\\
Mohammad Abdulaziz\inst{2,3}\orcidID{0000-0002-8244-518X} \and
Arnd Hartmanns\inst{1}\orcidID{0000-0003-3268-8674} \and\\
Peter Lammich\inst{1}\orcidID{0000-0003-3576-0504}
\thanks{%
This work was supported by
the European Union's Horizon 2020 research and innovation programme under Marie Sk{\l}odowska-Curie grant agreement
101008233 ({\scriptsize MISSION}),
the Interreg North Sea project {\scriptsize STORM\_SAFE},
NWO grant OCENW.KLEIN.311,
NWO VIDI grant VI.Vidi.223.110 ({\scriptsize TruSTy}),
NWO grant OCENW.M.21.291 ({\scriptsize VESPA})
and DFG grant 378803395 ({\scriptsize ConVeY}).
}%
}
\authorrunning{B. Kohlen et al.}
%
\institute{University of Twente, The Netherlands \email{b.kohlen@utwente.nl}\and
Technical University of Munich, Germany
\email{maximilian.schaeffeler@tum.de} \and
King's College London, United Kingdom
}

\maketitle              
\vspace{-1.5em}
\begin{abstract}
We present an efficiently executable, formally verified implementation of interval iteration for MDPs.
Our correctness proofs span the entire development from the high-level abstract semantics of MDPs to a low-level implementation in LLVM that is based on floating-point arithmetic.
We use the Isabelle/HOL proof assistant to verify convergence of our abstract definition of interval iteration and employ step-wise refinement to derive an efficient implementation in LLVM code.
To that end, we extend the Isabelle Refinement Framework with support for reasoning about floating-point arithmetic and directed rounding modes.
We experimentally demonstrate 
that the verified implementation is competitive with state-of-the-art tools for MDPs, while providing formal guarantees on the correctness of the results.
\end{abstract}
\section{Introduction}
\label{sec:introduction}

Probabilistic model checking (PMC)~\cite{Baier16,BaierAFK18} is a formal verification technique for randomized systems and algorithms like wireless communication protocols~\cite{KamaliK20}, network-on-chip (NoC) architectures~\cite{RobertsLHBRCZ21}, and reliability and performance models~\cite{BaierHHK10}.
Typical properties checked by means of PMC relate to \emph{reachability probabilities}:
What is the probability that a file will eventually be transmitted successfully~\cite{DArgenioJJL01}?
Is the probability that a NoC router's queue will overflow within $c$ clock cycles below $10^{-5}$?
What is the maintenance strategy that minimizes service outages within a given cost budget~\cite{RuijtersGNS16,RuijtersGDPS16}?
The system models that PMC is applied to are specified in higher-level modeling languages such as Modest~\cite{BohnenkampDHK06,HahnHHK13} or JANI~\cite{BuddeDHHJT17} with a formal semantics in terms of (extensions of) Markov chains and Markov decision processes (MDPs)~\cite{Bellman57,Puterman94}.

PMC delivers results with formal guarantees, typically that the computed and (unknown) true probabilities differ by at most a user-specified $\varepsilon$.
PMC is thus well-suited for the design and evaluation of safety- and performance-critical systems.
Over the past decade, however, we have witnessed several threats to the validity of PMC results.
First and foremost, the most-used PMC algorithm, value iteration (VI), was shown to be \emph{unsound} with an absolute- or relative-error stopping criterion, \ie produce arbitrarily wrong results for certain inputs~\cite{HaddadM14}.
Standard VI terminates whenever two successive iterates are sufficiently close in value.
However, at this point, the distance to the actual reachability probability may be arbitrarily off, no matter the error threshold \cite[Example 1]{HaddadM18}.
Several sound replacements for VI were subsequently developed~\cite{HaddadM18,HartmannsK20,QuatmannK18}, yet their soundness proofs have so far been \emph{pen-and-paper} style with room for human error.
For example, the pseudocode for the \emph{sound VI} algorithm as stated in~\cite{QuatmannK18} contains a subtle omission that only surfaces on 1 of the 78 models of the Quantitative Verification Benchmark Set (QVBS)~\cite{HartmannsKPQR19}.

This calls for \emph{formal specifications of the algorithms} accompanied by \emph{machine-checked correctness proofs}.
Even correct algorithms, however, may be incorrectly implemented in today's manually-coded PMC tools.
As a case in point, the implementation of the \emph{interval iteration} algorithm for expected rewards~\cite{BaierKLPW17} in the \mcsta model checker of the \toolset~\cite{HartmannsH14} diverges on some inputs.
We thus need \emph{correct-by-construction implementations}, too.

VI-based algorithms are iterative numeric approximation schemes that need to be implemented via machine-precision floating-point arithmetic to obtain acceptable performance~\cite{BuddeHKKPQTZ20,HartmannsJQW23}.
This introduces approximation and rounding errors that in turn may lead to incorrect Boolean outputs~\cite{WimmerKHB08}.
An efficient solution is to carefully use the directed rounding modes provided by standard IEEE 754 floating-point implementations on modern CPUs~\cite{Hartmanns22}, which, however, needs careful \emph{reasoning about floating-point errors and rounding} in all formal proofs and correctness-preserving implementation strategies.

\paragraph{Our contribution.}
We present a solution to all of the above challenges based on the interval iteration (II) algorithm~\cite{HaddadM18} for sound PMC on MDP models and the interactive theorem prover (ITP) \isabelle~\cite{NipkowPW02} with its Isabelle Refinement Framework (IRF)~\cite{LammichT12}:
\begin{itemize}
\item
We formalize (\ie model) II in \isabelle's logic and formally prove its correctness using \isabelle (\Cref{sec:interval_iteration}), making II the first sound PMC algorithm for MDPs with machine-checked correctness.
\item
We extend the IRF with support for floating-point arithmetic, including directed rounding modes (\Cref{sec:fp_arith}), introducing the first ITP-based algorithm refinement approach suitable for II and similar algorithms.
\item
Using the IRF, we refine the formalization of II into efficient LLVM bytecode (\Cref{sec:ii_refinement,sec:data_refinement}), delivering the first correct-by-construction implementation of a PMC algorithm.
\item 
In \Cref{sec:implementation}, we embed the code into \mcsta, a competitive probabilistic model checker.
We experimentally evaluate the performance using the QVBS, showing that the verified implementation is efficient.
Our formal proofs and the benchmark setup are available online.\footnote{\url{https://doi.org/10.4121/bf0fef24-4f0f-4de6-a58d-07b9ba601804}}
\end{itemize}

\paragraph{State-of-the-art: Verification of Algorithms for MDPs.}
A probabilistic model checker like \mcsta performs preprocessing and transformation steps for both correctness and performance.
Previously, the strongly connected component~\cite{HartmannsKL23} and maximal end component decomposition~\cite{HartmannsKL24} algorithms have been verified down to LLVM,
replacing their previous unverified implementations inside \mcsta by verified ones of comparable performance.
These were fully discrete graph algorithms, however, that neither required reasoning about numerical convergence in their correctness proofs nor floating-point arithmetic in their refinement to an efficient implementation.
With this work, we contribute an essential piece for the incremental replacement of unverified by verified algorithms for probabilistic reachability in \mcsta's MDP model checking core.

Other work relevant to our setting is the verification of iteration algorithms for MDPs: 
In Coq by Vajjha et al.~\cite{VajjhaSTPF21} and in Isabelle/HOL by Sch\"affeler and Abdulaziz~\cite{valIterIsabelle,ApproxPolIterIsabelle}.
These works contribute formalizations of the classical version of value iteration and policy iteration that optimize the expected discounted values, and a modified policy iteration algorithm for solving large, factored MDPs.
We note that only Sch\"affeler and Abdulaziz~\cite{valIterIsabelle,ApproxPolIterIsabelle} also verified practical implementations.
However, since their implementations used infinite-precision arithmetic, they could not compete with state-of-the-art floating-point implementations.
Thus, the work we present here is the first, to our knowledge, where a full formal mathematical analysis of an algorithm, involving heavy usage of a formal mathematical probabilities library, is performed and a competitive floating-point implementation is also verified.
Furthermore, from a formalization-methodology perspective, we note that the correctness argument of II involves a substantial element of graph-theoretic reasoning, in addition to the reasoning about fixed points that is present in II and other verified MDP algorithms.
This includes reasoning about connected components, acyclicity, and levels in a directed acyclic graph (DAG),
further complicating II's verification compared to other verified iteration algorithms.

\paragraph{State-of-the-art: Verification of Floating-Point Algorithms.}
It is widely recognized as a problem that floating-point implementations often deviate from the mathematical models of the underlying algorithms.
Bugs with potentially serious consequences were noted in the hardware and aerospace industry~\cite{floatingPointHOLLight,moscatoFloatingPoint2017}.
Due to the complexity of floating-point algorithms' behavior, and the failure of testing to reliably catch bugs in those algorithms, there is a long tradition of applying formal methods to the verification of floating-point algorithms.
This was done in verification systems like Z~\cite{floatingPointZ}, HOL Light~\cite{floatingPointIntel,floatingPointHOLLight}, PVS~\cite{boldo2006high,miner1995defining}, and Coq~\cite{flocq,de2010certifying}.
Most of that previous work, however, focused on proving correctness of basic algorithms implemented in floating-point arithmetic, such as foundational linear algebra operations~\cite{KellisonATB23}.
In contrast, we aim to do the correctness proofs on algorithms using real numbers, which we implement as floating-point numbers with directed rounding.
This keeps our correctness proofs manageable while preserving interesting properties, even for complex programs.

A related line of work aims to prove correctness by providing error bounds.
Tools like PRECiSA~\cite{moscatoFP2025}, VCFloat2~\cite{AppelK24}, FPTaylor~\cite{SolovyevBBJRG19}, Real2Float~\cite{MagronCD17}, and Fluctuat~\cite{GoubaultP11} analyze the floating-point error propagation.
They focus on determining the worst-case roundoff error.
While more expressive than our approach, these tools have limited to no support for programs with complex control flow like the nested loops in the implementation of II.

The static analysis tool Astr\'ee~\cite{CousotCFMMMR05} is used in the aviation and automotive industry to check the absence of runtime errors.
While it supports floating-point arithmetic, it cannot verify general correctness properties.
Frama-C~\cite{KirchnerKPSY15} has similar functionality but also features deductive verification.
However, the properties supported are limited to, \eg, proving that outputs lie in a given interval~\cite{KosmatovPS24}.
The situation is similar with other deductive verifiers such as KeY~\cite{KeYBook16}, which can verify the absence of exceptional floating-point values like \emph{NaN} and \emph{infinity}~\cite{AbbasiSDUA21}.

\section{Preliminaries}
\label{sec:prelims}

We now present the necessary background for the rest of the paper: we introduce \isa and the Isabelle Refinement Framework, followed by IEEE 754 floating-point numbers and Markov decision processes in \isabelle.

\subsection{Isabelle/HOL}
An \emph{interactive theorem prover} (ITP) is a program that implements a formal mathematical system in which a user writes definitions and theorem statements, and constructs proofs from a set of axioms.
To prove a theorem in an ITP, the user provides high-level steps, and the ITP fills in the details at the axiom-level.

We perform our formalization using the ITP \isabelle~\cite{NipkowPW02}, which is a proof assistant for \emph{higher-order logic} (HOL).
Roughly speaking, HOL can be seen as a combination of functional programming with logic.
\isa is designed to be highly trustworthy: a small, trusted kernel implements the inference rules of the logic. 
Outside the kernel, a large set of tools implement proof automation and high-level concepts like algebraic data types.
Bugs in these tools cannot lead to inconsistent theorems being proved, 
as the kernel refuses flawed proofs.

We aim to represent our formalization as faithfully as possible, but we have optimized the presentation for readability.
The notation in \isabelle is similar to functional programming languages like ML or Haskell mixed with mathematical notation.
Function application is written as juxtaposition: we write $f~x_1~\ldots~x_n$
instead of the standard notation $f(x_1,~\ldots~,x_n)$.
Recursive functions are defined using the \is{fun} keyword and pattern matching.
For partial functions, we use the notation \is{f = (\<lambda>x \<in> X. g x)}, to explicitly restrict the domain of the function to \is{X}.
Where required, we annotate types as \is{x :: type}.

\isabelle provides a keyword \is{locale} to define a named context with assumptions, \eg an MDP with well-formedness assumptions~\cite{Ballarin03}.
Locales can be interpreted and extended in different contexts, \eg, a locale for MDPs can be instantiated for a specific MDP, which yields all theorems from that locale.

\subsection{Isabelle Refinement Framework}
Our verification of II spans the mathematical foundations of MDPs, the implementation of optimized algorithms and data structures, and the low-level LLVM intermediate language~\cite{LattnerA04}.
To keep the verification effort manageable, we use a stepwise refinement approach: starting with an abstract specification,
we incrementally add implementation details, 
proving that each addition preserves correctness,
\eg, computing a fixed point by iteration, or implementing MDPs by a sparse-matrix data structure.
The former specifies the control flow of a program, but the data type remains the same.
The latter we call \emph{data refinement}.

This approach is supported by the Isabelle Refinement Framework (IRF)~\cite{LammichT12}.
In the IRF, we define algorithms in the \emph{nondeterministic result (nres) monad}, where a program either fails or produces a set of results.
The notation $a$ \refcomp $c$ denotes that every (non-deterministic) output of abstract program $a$ is related to an output of concrete program $c$ via the \emph{refinement relation} $R$.
In other words, $c$ is an implementation of $a$.
If a refinement step does not involve data refinement, then we use $\refcomp[]\ \colonequals\ \refcomp[\Rid]$ where \Rid is the identity relation.

We target LLVM for our code generation, for which the IRF provides semantics, so that we can perform refinements to LLVM in a verified way.
LLVM is an intermediate representation, meaning that programming languages like C and Rust can be compiled to LLVM.
The LLVM compiler framework can then be used to compile to bytecode and apply optimizations.
In our formalization, the refinement chain starts with a high-level specification of the algorithm and ends with an efficient LLVM program.
The \sepref~\cite{Lammich19} tool can automatically refine a program to LLVM and prove a refinement theorem, given that we have provided compatible data structures (\sepref can automatically refine HOL lists to arrays, nats to 64-bit integers, etc.).
Due to transitivity of refinement, the LLVM program satisfies the specification.

\subsection{Floating-Point Arithmetic}

Our work builds on a formalization of the IEEE~754 floating-point standard in \isabelle~\cite{Yu13}.
This library provides a generic type \is{(e,f) float} resembling scientific notation: \is$e$ is the number of bits for the \emph{exponent},
and \is$f$ is the number of bits for the \emph{fraction} (also known as mantissa).
We use the type \is{double = (11,52) float} which is the standard double precision format.

The floating-point format contains positive and negative numbers, as well as designated special values for $\pm \infty$ and \emph{not a number (\nan)}.
The function \is{valof :: (e,f) float -> ereal} maps non-\nan floating-point numbers to \emph{extended real numbers}, i.e. $\mathbb{R}~\dot\cup~\{-\infty,+\infty\}$.
Finally, the formalization provides standard floating-point instructions like addition, multiplication, and comparisons as well as intuitive predicates to identify special cases (\eg \is{is_nan}).

\subsection{Markov Decision Processes}
\label{sec:mdps_isabelle}

\emph{Markov decision processes} (MDPs) are widely used to model probabilistic systems with nondeterministic choices~\cite{Puterman94}, \eg, in PMC, planning, operations research and reinforcement learning~\cite{SuttonB98,BaierK08}.
Intuitively, an \emph{agent} interacts with an environment by choosing \emph{actions} that, together with random elements, influence the \emph{state} of the system.
The agent has an \emph{objective}, \eg, to reach certain states, and aims to choose actions that optimize the probability to achieve the objective.
The important concepts introduced in this section are illustrated in \Cref{ex:mdp}. 

Formally, a finite MDP is a pair $\mdp = (\states, \kernel)$ where
$\states$ is a finite, non-empty set of \emph{states}, and $\kernel : \states \to 2^{\pmf(\states)}$ is the transition kernel.
It maps every state to a finite, non-empty set of \emph{actions} in the form of transition probabilities.
$\pmf(\states)$ denotes the set of probability measures on $\states$, \ie functions $p : \states \to [ 0, 1 ]$ where $\sum_{s \in \states} p(s) = 1$.
Furthermore $\kernel$ is closed under $\states$: Actions from $\states$ lead to $\states$.

Our formalization of II extends the \emph{Markov Models}~\cite{HolzlN12,Holzl17} formalization from \emph{archive of formal proofs} (AFP)---a collection of \isa libraries.
MDPs are modeled with a generic type \is{'s mdpc} and a locale \is{Finite_MDP} that, in combination, contain the states, the transition kernel and well-formedness conditions~(\Cref{lst:mdp}).
In the following, we abbreviate the projections \is{states M} and \is{actions M} as $S$ and $K$. 
The type of the states is \is{'s}, and the type of probability distributions over \is{'s} is \is{'s pmf}.
For a distribution \is{p :: 's pmf}, \is{set_pmf p} denotes its support, i.e. the set of states with non-zero probability.

\begin{isalocale}[label=lst:mdp]
locale Finite_MDP = %\hfill (\Cref{lst:mdp})%
  fixes M :: "'s mdpc" and S and K
  defines "S = states M" and "K = actions M"
  assumes "S \<noteq> {}" and "finite S" and "\<forall>s. K s \<noteq> {}" and "\<forall>s \<in> S. finite (K s)"
  assumes "\<forall>s \<in> S. \<Union>a \<in> K s. set_pmf a \<subseteq> S"
\end{isalocale}

\begin{wrapfigure}{r}{0.4\textwidth}
\centering
\vspace{-20pt}
	\begin{tikzpicture}[on grid,scale=1, node distance=2.7cm, nodestyle/.style={draw,circle,minimum size=0.75cm},baseline=(s0)] 
    
	   		\node [nodestyle] (s0) at (0,0) [on grid] {$0$};
			\node [nodestyle] (s1) [on grid, right of= s0] {$1$};
			\node [nodestyle] (s2) [on grid, below = 1.9 of s0] {$2$};
			\node [nodestyle] (s3) [on grid, right of= s2] {$3$};
			
			\draw (s0) edge[->, bend right] node[pos=0.20, left] {$\alpha$} node[draw, circle,  inner sep=2pt, fill] (s0b) {} node [pos=0.8, left] {$1$} (s2);
			
			\draw (s0) edge[->, bend right=40] node[pos=0.25, above] {$\beta$} node[draw, circle,  inner sep=2pt, fill] (s0a) {} node [pos=0.75, above] {$0.1$} (s1);
			\draw (s0a) edge[->, bend left=60,out=120] node[pos=0.6, below] {$0.9$} (s0);
			
			\draw (s1) edge[->, bend right] node[pos=0.25, below] {$\gamma$} node[draw, circle,  inner sep=2pt, fill] (s1a) {} node [pos=0.75, below] {$1$} (s0);			
			
			\draw (s2) edge[->, bend right] node[pos=0.25, above] {$\delta$} node[draw, circle,  inner sep=2pt, pos=0.65, fill] (s2a) {} node [pos=0.85, right] {$0.5$} (s1);
			\draw (s2a) edge[->, bend left] node[pos=0.35, above right, yshift=-4pt] {$0.5$} (s3);

			\draw (s3) edge[->, loop left] node[pos=0.2, below] {$\epsilon$} node[draw, circle,  inner sep=2pt, fill, below=-1mm] (s3a) {} node [pos=0.8, above] {$1$} (s3);
		\end{tikzpicture}
	\caption{A simple MDP with four states and five actions.}
	\label{fig:mdp}
\vspace{-10pt}
\end{wrapfigure}

At every state, the agent chooses an action based on the current state and the history of visited states.
The agent's choices are captured by a \emph{scheduler} (aka policy, strategy, or adversary), which is a function that maps histories to actions.
The MDP formalization works with \emph{configurations}, which are pairs of states and strategies.
A configuration is \emph{valid} if the strategy selects only enabled actions and the state of the configuration is in \is{S}.
\is{valid_cfg} denotes the set of all valid configurations.
Given a configuration and an MDP, the probability space of infinite traces \is{T cfg} is constructed from the induced Markov chain, where each state is a configuration.

MDP subcomponents play an important role in the analysis of II.
A \emph{sub-MDP} $\mdp' = (\states', \kernel')$ consists of a subset of states $\states' \subseteq \states$ and a restricted kernel $\kernel'$ where $\forall s \in \states'.~\kernel'(s) \subseteq \kernel(s)$. 
A sub-MDP is \emph{strongly connected} if all states are connected via a sequence of actions.
A closed, strongly connected sub-MDP is an \emph{end component} (EC).
A \emph{maximal end component} (MEC) is an EC that is not a sub-MDP of another EC.
Finally, \emph{trivial MECs} are MECs with one state and no actions, and \emph{bottom MECs} are MECs without an exit.

\paragraph{Reachability.}
In our PMC setting, the objective is to minimize or maximize the long-term reachability probabilities of a set of target states $U \subseteq S$.
The value function \is{\v :: 's => real} gives the probability of reaching \is{U} in the Markov chain induced by the configuration \is{cfg}. 
Minimal and maximal reachability probabilities are denoted by \is{\n} and \is{\p} respectively. 
\is{\n} is defined as the infimum of \is{\v} over all valid configurations, \is{\p} is defined using the supremum.
We also introduce the \emph{Bellman optimality operators} \is{F_inf} and \is{F_sup} (\Cref{lst:F_def}, \is{F_sup} omitted). For a state \is{s \in S} and a value vector \is{v}, \is{F_inf v s} denotes the minimal expected value of \is{v} after taking a single step from \is{s}.
The symbol \is{\<Sqinter>} denotes the infimum.
\begin{isadefinition}[label=lst:F_def]
  definition F_inf v = (\<lambda>s \<in> S. %\hfill (\Cref{lst:F_def})%
    if s \<in> U then 1
    else \INFa \<in> K s. \<Sum>t \<in> set_pmf a. v t $\cdot$ pmf a t)" 
\end{isadefinition}

The \emph{least fixed point} (\is{lfp}) of \is{F_inf} is \is{\n}.
In other words, repeatedly applying \is{F_inf} to a lower bound of \is{\n} converges to \is{\n} in the limit.
For II, we preprocess the MDP such that the \emph{greatest fixed point} (\is{gfp}) of \is{F_inf} also equals \is{\n}, and then
iterate \is{F_inf} on both a lower and an upper bound until they closely approximate \is{\n}.
The same holds for \is{F_sup} and \is{\p}.

\begin{example}\label{ex:mdp}
    \Cref{fig:mdp} shows an MDP with four states, $S = \{ 0, 1, 2, 3 \}$.
    The outgoing transitions from each state represent the actions in the MDP.
    Each transition leads to a black dot and branches into the successor states with corresponding probabilities.
    For example in state $0$, $K~0 = \{ \alpha, \beta \}$. 
    The agent can choose action $\alpha$ to move to state $2$, or $\beta$ to have a $10\%$ chance to move to state $1$.

    Let the target states $\target = \{ 3 \}$.
    The reachability probabilities are \is{\n $2$ = $0.5$} and \is{\p $2$ = $1$}.
    The MDP has a single bottom MEC $\{ 3 \}$ with action $\{\epsilon\}$, and a single trivial MEC $\{ 2 \}$.
    The states $\{ 0, 1 \}$ form a MEC with actions $\beta$ and $\gamma$.
\end{example}

\section{Interval Iteration in Isabelle/HOL}
\label{sec:interval_iteration}

The \emph{interval iteration (II)} algorithm for MDPs is an iterative solution method for reachability problems based on value iteration.
In contrast to standard value iteration applied to PMC, II provides a simple and sound stopping criterion.

We present our \isabelle formalization of definitions and correctness proofs for II and its preprocessing routines.
Our formalization is based on the proofs in \cite{HaddadM18}: We highlight the challenges encountered during formalization and point out differences in our formal proofs to their pen-and-paper equivalents.
In particular, we present a more elegant and much more precise proof of [22, Proposition 3].
Moreover, we simplify the definitions of the preprocessing steps significantly.
In the following, all statements prefixed with \is{lemma} or \is{theorem} are formally verified in \isabelle.
All theorems and definitions for \is{\p} that are analogous to the ones for \is{\n} are omitted here for brevity.

\subsection{The Interval Iteration Algorithm}
The idea of the II algorithm is to start with a lower and an upper bound on the true reachability probabilities and then iterate the Bellman optimality operator \is{F_inf} (or \is{F_sup}) on both.
Since the optimality operators are monotone, both sequences converge to a fixed point.
On arbitrary MDPs, these fixed points are not necessarily the same.
However, if the MDP is preprocessed to only contain MECs that are trivial or bottom MECs, both fixed points are equal to the optimal reachability probabilities.

For now, assume an arbitrary MDP with a single target state \is{Inr True} and an avoid state \is{Inr False}, that are both sinks.
As the initial lower bound \is{lb_0}, we take the function that assigns $1$ to \is{Inr True} and $0$ to all other states.
The initial upper bound \is{ub_0} assigns $0$ to \is{Inr False} and $1$ to all other states.
The II algorithm computes the sequences \is{lb_n n} and \is{ub_n n}, defined as the $n$-fold application of \is{F_inf} to either \is{lb_0} or \is{ub_0}:

\begin{isadefinition}[label=lst:iters]
definition \<open>lb_n n = (F_inf)^n lb_0\<close>  and \<open> ub_n n = (F_inf)^n ub_0\<close> %\hfill (\Cref{lst:iters})%
\end{isadefinition}
It is an immediate consequence of the monotonicity of the Bellman optimality operators that the lower (upper) bounds are monotonically increasing (decreasing).
Clearly, \is{lb_0} is a lower bound and \is{ub_0} is an upper bound of \is{\n}. 
Additionally, we formally derive that the Bellman optimality operators preserve upper and lower bounds in \Cref{lst:bounds_preserved}.
These two properties of the abstract II algorithm are required for the refinement proof in \Cref{sec:fp_refinement}.

\begin{isalemma}[label=lst:bounds_preserved]
lemma  assumes \<open>v \<le> \n\<close>  shows \<open>F_inf v \<le> \n\<close>%\hfill (\Cref{lst:bounds_preserved})%
lemma  assumes \<open>v \<ge> \n\<close>  shows \<open>F_inf v \<ge> \n\<close>
\end{isalemma}%

\subsection{Reduced MDPs}
\label{sec:reduction}

II is only guaranteed to converge if the MDP only contains trivial or bottom MECs.
We therefore need to preprocess the MDP before applying II.
The preprocessing steps differ for \is{\n} and \is{\p}, they are called \emph{min-reduction} and \emph{max-reduction}.
In a first step, we extend the existing MDP formalization~\cite{Holzl17} with \emph{strongly connected components}~(SCCs) and bottom MECs (\Cref{lst:bmec}).
The states of an MDP that form trivial or bottom MECs are called \is{trivials} or \is{bottoms} respectively.
We follow the original presentation of II~\cite{HaddadM18} and call an MDP \emph{reduced} if all of its MECs are either trivial or bottom MECs.

\begin{isadefinition}[label=lst:bmec]
definition "bmec M b = %\hfill (\Cref{lst:bmec})% 
    mec M b \<and> (\<forall>s \<in> states b. actions b s = K s)"
\end{isadefinition}

\paragraph{Min-Reduction.}
The min-reduction for MDPs transforms an arbitrary MDP into a reduced MDP whilst \is{\n} remains unchanged.
Observe that we may assign a minimal reachability of $0$ to all non-trivial MECs (except $s_+$):
There exists a strategy that stays in the MEC forever and therefore never reaches $s_+$.
Hence, all such MECs may be collapsed into a single absorbing state $s_-$.
To formalize this transformation, we first define a function \is{min_red}~(\Cref{lst:red}) to map single states, and then apply it to the MDP \is{M} to obtain the reduced MDP \is{M_red}~(\Cref{lst:M_red}).
Our formal definition is a substantial simplification compared to the pen-and-paper version~\cite[Def. 4]{HaddadM18}.

The function \is{map_mdpc} (defined by H\"olzl~\cite{HolzlN12}) applies a function to every state of an MDP.
If the function merges states, \is{map_mdpc} merges the action sets.
Finally, \is{fix_loop Inr False} replaces the actions at \is{Inr False} with a single self-loop.
\Cref{fig:minmdp} displays the min-reduced version of the MDP from \Cref{fig:mdp}.

\begin{isadefinition}[label=lst:red]
definition min_red s = if s \<in> trivials M \<union> {Inr True} then s else Inr False\<close> %\hfill (\Cref{lst:red})%
\end{isadefinition}
\vspace{-0.5em}
\begin{isadefinition}[label=lst:M_red]
definition \<open>M_red = fix_loop Inr False (map_mdpc min_red M)\<close> %\hfill (\Cref{lst:M_red})%
\end{isadefinition}

\begin{figure}[t]
  \begin{minipage}{.45\textwidth}
    \centering
    	\begin{tikzpicture}[on grid,scale=1, node distance=2.7cm, nodestyle/.style={draw,circle,minimum size=0.75cm},baseline=(s0)] 
    
	   		\node [nodestyle] (s0) at (0,0) [on grid] {$s_-$};
			\node [nodestyle] (s3) [on grid, below = 1.9 of s0] {$s_+$};
			\node [nodestyle] (s2) [on grid, left of=s3] {$2$};
			
			\draw (s2) edge[->, bend right] node[pos=0.25, above] {$\delta$} node[draw, circle,  inner sep=2pt, pos=0.65, fill] (s2a) {} node [pos=0.85, left] {$0.5$} (s0);
			\draw (s2a) edge[->, bend left] node[pos=0.35, above right, yshift=-4pt] {$0.5$} (s3);

			\draw (s3) edge[->, loop left] node[pos=0.05, left] {$$} node[draw, circle,  inner sep=2pt, fill, below=-1mm] (s3a) {} node [pos=0.8, above] {$1$} (s3);
			\draw (s0) edge[->, loop left] node[pos=0.05, left] {$$} node[draw, circle,  inner sep=2pt, fill, below=-1mm] (s3b) {} node [pos=0.8, above] {$1$} (s3);
		\end{tikzpicture}
	  \caption{Min-reduced MDP derived from the MDP in \Cref{fig:mdp}.}
	  \label{fig:minmdp}
  \end{minipage}%
  \hfill
  \begin{minipage}{.45\textwidth}
    \centering
    	\begin{tikzpicture}[on grid,scale=1, node distance=2.7cm, nodestyle/.style={draw,circle,minimum size=0.75cm},baseline=(s0)] 
    
	   		\node [nodestyle] (s0) at (0,0) [on grid] {$01$};
			\node [nodestyle] (s2) [on grid, below = 1.9 of s0] {$2$};
			\node [nodestyle] (s3) [on grid, right of= s2] {$s_+$};
			\node [nodestyle] (s1) [on grid, right of= s0] {$s_-$};
			
			\draw (s0) edge[->, bend right] node[pos=0.20, left] {$\alpha$} node[draw, circle,  inner sep=2pt, fill] (s0b) {} node [pos=0.8, left] {$1$} (s2);

			\draw (s2) edge[->, bend right] node[pos=0.25, left] {$\delta$} node[draw, circle,  inner sep=2pt, fill] (s2a) {} node [pos=0.8, right] {$0.5$} (s0);
			\draw (s2a) edge[->, bend left] node[pos=0.4, above] {$0.5$} (s3);

			\draw (s3) edge[->, loop left] node[pos=0.05, left] {$$} node[draw, circle,  inner sep=2pt, fill, below=-1mm] (s3a) {} node [pos=0.8, above] {$1$} (s3);
			\draw (s1) edge[->, loop left] node[pos=0.05, left] {$$} node[draw, circle,  inner sep=2pt, fill, below=-1mm] (s1a) {} node [pos=0.8, above] {$1$} (s1);
		\end{tikzpicture}
	  \caption{Max-reduced MDP derived from the MDP in \Cref{fig:mdp}.}
    \label{fig:maxmdp}
  \end{minipage}
  \end{figure}

Our formal proof of the fact that \is{M_red} is in fact a reduced MDP follows the original proof~\cite{HaddadM14}.
Next, we also need to show that the transformation preserves \is{\n}. 
To distinguish the reachability probabilities of the original and the reduced MDP, we use the notation \is{\n} for the original MDP and \is{M_red.\n} for the reduced MDP.
Our correctness proof of the transformation is based on the fact that min-reduction preserves the finite-horizon probabilities \is{n_fin n} (\Cref{lst:n_fin_eq_red}), \ie, the reachability probability in $n$ steps.
Now, the main claim (\Cref{lst:n_eq_red}, \cite[Proposition~3]{HaddadM18}) is a direct consequence.
Note that our proof is simpler and more precise than the original: 
Haddad and Monmege~\cite{HaddadM18} only claim without details that for every strategy in the reduced MDP, there exists a strategy in the original MDP with the same \is{\n} and vice versa.

\begin{isalemma} [label=lst:n_fin_eq_red]
lemma  assumes \<open>s \<in> S\<close>  shows \<open>n_fin n s = M_red.n_fin n (min_red s)\<close>%\hfill (\Cref{lst:n_fin_eq_red})%
\end{isalemma}
\vspace{-0.5em}
\begin{isatheorem}[label=lst:n_eq_red]
theorem  assumes "s \<in> S"  shows "\n s = M_red.\n (min_red s)" %\hfill (\Cref{lst:n_eq_red})%
\end{isatheorem}

\paragraph{Max-Reduction.}
The \is{\p} case can be handled similarly with a max-reduction. 
Yet the procedure is more involved, as not all non-trivial MECs can be collapsed while preserving \is{\p}---a maximizing strategy might choose to leave a non-trivial MEC.
We can, however, first collapse each MEC into a single state to obtain an MDP \is{M_MEC}.
This step keeps the reachability probabilities unchanged.
In a second step, we map the bottom MECs to \is{Inr False}.
Finally, we remove self-loops at all states except \is{Inr True} and \is{Inr False}.
Our formalization decomposes max-reduction~\cite[Def. 5]{HaddadM18} into multiple steps, which in turn simplifies both definitions and proofs.

Collapsing the MECs into a single state is performed by the function \is{the_mec}, the transformation preserves \is{\p} (\Cref{lst:p_eq_mec}).
Our proof resembles the proof of \cite[Theorem 3.8]{Alfaro97}, however we have to work around the fact that the MDP formalization~\cite{Holzl17} only supports deterministic policies.
Note that every state of \is{M_MEC} now forms its own MEC.
The correctness proof of the second phase of the reduction
is similar to min-reduction.
See \Cref{fig:maxmdp} for the max-reduced version of the MDP from \Cref{fig:mdp}.

\begin{isatheorem}[label=lst:p_eq_mec]
theorem %\hfill (\Cref{lst:p_eq_mec})%
    assumes "s \<in> S"
    shows "\p s = M_MEC.\p (the_mec M s)"
\end{isatheorem}
\vspace{-1em}
\begin{proof}
($\leq$) As collapsing MECs only shortens paths in the MDP, the proof proceeds similarly to the one for min-reduction via finite-horizon probabilities.

($\geq$) 
Consider an optimal strategy $\pol_{\mathit{MEC}}$ in \is{M_MEC}, we need to show that there exists a strategy in \is{M} with the same reachability probability.
Every MEC $m$ of $M$ contains a state $s^\pi_m$, where the action selected by $\pol_{\mathit{MEC}}$ is enabled.
Moreover, within a MEC, we can obtain a deterministic, memoryless strategy $\pol_m$ that reaches this state with probability $1$.
Thus, we can construct a strategy in \is{M} that behaves like $\pi_m$ within each MEC until $s^\pi_m$ is reached and then follows $\pol_{\mathit{MEC}}$.
The reachability probability of this strategy in \is{M} is the same as the one achieved by $\pol_{\mathit{MEC}}$ in \is{M_MEC}.
\end{proof}

\subsection {Reachability in Reduced MDPs}
From now on, we assume that we are working with a reduced, finite MDP \is{M},
where each state is either a trivial or a bottom MEC.
We show that in such an MDP, over time, any strategy reaches a bottom MEC almost surely.
This is the key property that will then allow us to prove the convergence of II.

\paragraph{Level Graph.}
First, we build a level graph of the MDP, starting at the bottom MECs (\Cref{lst:G}).
At level $n + 1$, we add all those states where every action has a successor on level $n$ or below.
We define \is{I}\, to be the greatest non-empty level of the level graph \is{G}, so \is{I}\, is the number of steps that allows us to reach a bottom MEC from every state.
We formally show that \is{G} has the desired properties, i.e. it is acyclic and contains every state at exactly one level.
The proofs in \isabelle require substantial reasoning about graph-theoretic properties, \eg, we need to show that every MDP contains a bottom MEC.

\begin{isadefinition}[label=lst:G]
fun G where %\hfill (\Cref{lst:G})%
  "G 0 = bottoms M"
  "G (Suc n) = let G_le = \<Union>i \<le> n. G i in
    {s \<in> S \<setminus> G_le. \<forall>a \<in> K s. G_le \<inter> a \<noteq> {}}"
\end{isadefinition}

\paragraph{Reachability of BMECs.}
We now show that, intuitively, every strategy eventually descends through the levels of \is{G}.
The rate at which a bottom MEC is encountered is bounded in terms of \is{min_prob}, the smallest probability of any transition in the MDP.
At every step, the probability of descending a level with respect to \is{G} is at least \is{min_prob}.
Hence, we can show that for any valid configuration, the probability to reach the bottom MECs in $\levels$ steps is no less than \is{min_prob^I} (\Cref{lst:v_fin_I_ge}).

\begin{isalemma}[label=lst:v_fin_I_ge]
lemma  assumes \<open>cfg \<in> valid_cfg\<close>  shows \<open>min_prob^I \<le> v_fin I\<close>%\hfill (\Cref{lst:v_fin_I_ge})%
\end{isalemma}

The value \is{v_fin n} denotes the finite-horizon reachability probability of the bottom MECs in $n$ steps under configuration \is{cfg}.
Note that the lemma was originally stated for safety instead of reachability problems.
We transform it using the well-known equivalence
$\prfin{U}{\textit{cfg}}{n} = 1 - \safetyfin{\lnot U}{\textit{cfg}}{n}$.
For $n$ multiples of $\levels$, we obtain the stronger lower bound \is{1 - (1 - min_prob^I)^n} (\Cref{lst:v_fin_Ii_ge}, \cite[Proposition~1]{HaddadM18}).
As $n$ increases, \is{(1 - min_prob^I)^n} converges to $0$ and thus \is{v_fin nI} tends towards $1$.
Since we chose \is{cfg} arbitrarily, we almost surely reach a bottom MEC in the limit.

\begin{isatheorem}[label=lst:v_fin_Ii_ge]
theorem  assumes \<open>cfg \<in> valid_cfg\<close> shows \<open>1 - (1 - min_prob^I)^n \<le> v_fin nI \<close>%\hfill (\Cref{lst:v_fin_Ii_ge})%
\end{isatheorem}

\subsection{Convergence of Interval Iteration}
We assume a special form of reduced MDPs, where the only bottom MECs are the target state $\reach$ and avoid state $\avoid$, that are both absorbing (Locale~\ref{lst:reduced_target}).
The reduced MDPs from \Cref{sec:reduction} are instances of this locale.

\begin{isalocale}[label=lst:reduced_target]
locale MDP_Reach = Finite_MDP M + %\hfill (\Cref{lst:reduced_target})%
  assumes Inr False \<in> S and Inr True \<in> S and 
    \<forall>s \<in> S \<setminus> {Inr True, Inr False}. s \<in> trivials M\<close> and
    K Inr False = {return_pmf Inr False} and K Inr True = {return_pmf Inr True}
\end{isalocale}

Towards a convergence proof of II, we show that the lower and upper bound sequences are equal to the finite-horizon reachability probabilities towards \is{Inr True} and \is{Inr False} respectively (\Cref{isa:bds_eq_fin}, \cite[Lemma~4]{HaddadM18}).
For improved clarity, we indicate target sets explicitly in this section.

\begin{isalemma} [label=isa:bds_eq_fin]
  lemma %\hfill (\Cref{isa:bds_eq_fin})%
    assumes "s \<in> S" 
    shows \<open>lb_n n s = n_fin {Inr True} n s\<close>  and  \<open>ub_n n s = 1 - p_fin {Inr False} n s\<close>
  \end{isalemma}
Combining the above result with \Cref{lst:v_fin_Ii_ge}, we can bound the distance between the sequences~(\Cref{lst:bounds_diff}).
Note that convergence is in general only guaranteed if all computations are carried out with arbitrary precision arithmetic.
In a floating-point setting, the convergence to a unique fixed point is not guaranteed.
Still, this theoretical result motivates the usage of the II algorithm to optimally solve reachability problems on MDPs.
In practice, on most instances the algorithm converges much faster than the theoretical bound suggests (see \Cref{sec:implementation} for experimental results).

Finally, the theorem does not apply if all probabilities in the MDP are equal to $1$, \ie no branching occurs after action selection~\cite{HaddadM18}.
In this case, the MDP is deterministic and is better solved with qualitative solution methods.

\begin{isatheorem}[label=lst:bounds_diff]
theorem %\hfill (\Cref{lst:bounds_diff})%
  assumes \<open>s \<in> S\<close>  and  \<open>\eps > 0\<close>  and  \<open>min_prob \<noteq> 1\<close>  and  \<open>n \<ge> \<lceil>%$\log_{(1 - \eta^I)} \epsilon$%\<rceil> * I\<close>
  shows \<open>ub_n n s - lb_n n s \<le> \eps\<close>
\end{isatheorem}
\vspace{-1em}\begin{proof}
As a first step, we show for all \is{i} that 
\begin{lstlisting}
  ub_n iI s - lb_n iI s = 1 - p_fin {Inr False} iI s - n_fin {Inr True} iI s %\hfill (Lemma \ref{isa:bds_eq_fin})%
    \<le> 1 - (n_fin {Inr False} iI s + n_fin {Inr True} iI s) %\hfill (%n_fin \<le> p_fin%)%
    = 1 - n_fin {Inr False, Inr True} iI s %\hfill (Disjoint events)%
    \<le> (1 - min_prob^I)^i. %\hfill (\Cref{lst:v_fin_Ii_ge})%
\end{lstlisting}
Set $i = \lceil \log_{(1 - \eta^I)} \epsilon \rceil$ and the theorem follows from monotonicity.
\end{proof}

\section{Refinement using Floating-Point Arithmetic}
\label{sec:fp_refinement}

In the next step, we use the IRF to refine II to an efficient LLVM implementation, refining real numbers to IEEE~754 double precision floating-point numbers~(floats).
While our framework can be used to refine to floats of any precision, we focus on the widely-adopted double precision variant here for smaller rounding errors than single-precision floats.

To avoid complex reasoning about error bounds, we propose a safe rounding approach that refines reals to \emph{upper bounding} or \emph{lower bounding} floats.
As the name suggests, upper bounding floats are always greater than or equal to the real value they refine, while lower bounding floats are always less than or equal.

We first introduce the \sepref tool for refinement to LLVM, after which we present our extension of the IRF and \sepref for floats. 
This new reasoning infrastructure for floating-point implementations serves as the framework to obtain correct-by-construction LLVM code for II.

\subsection{The Sepref Tool}
\label{sec:sepref}
We use \sepref~\cite{Lammich19} to automatically refine algorithms to LLVM programs.
The \sepref tool provides a library of verified, reusable data structures.
Since LLVM maintains a \emph{heap}, we need to be able to deal with memory allocation and disposal.
We use \emph{assertions} from \emph{separation logic}~\cite{Reynolds02} which are refinement relations extended with a heap.
Formally \is{A :: 'a => 'c => 'h => bool} where \is{'a} and \is{'c} is the abstract and concrete data and \is{'h} is the heap.
The assertion holds if \is{a::'a} refines \is{c::'c} and \is{c} is encoded in heap \is{h::'h}.
\sepref uses separation logic to ensure memory correctness, \eg, correct disposal and preventing dangling pointers.

\begin{example}
\label{ex:sepref}

The following example demonstrates how to employ \sepref for automatic refinements to LLVM code.

\begin{lstlisting}[numbers=left, label=lst:sepref_ex]
	definition ls$\sunderscore$app x xs = xs @ [x] and half$\sunderscore$nat n = n div 2%\label{exl:append}%
	definition app$\sunderscore$half n xs = do { let n' = half$\sunderscore$nat n; return ls$\sunderscore$app n' xs } %\label{exl:add}%
	lemma (rshift1, half$\sunderscore$nat) :: \shiftleft$\Asize$ \<rightarrow> $\Asize$ %\label{exl:nref}%\shiftleft
	lemma (arl$\sunderscore$app, ls$\sunderscore$app) :: [\<lambda>n xs. length xs < $2^{63}\text{\scriptsize{-}}1$] \shiftleft$\Asize$ \shiftleft\<rightarrow>\shiftleft $A_{\mathit{arl}}^d$ \shiftleft\<rightarrow>\shiftleft $A_{\mathit{arl}}$ %\label{exl:aref}%
	lemma (arl$\sunderscore$app$\sunderscore$half, app$\sunderscore$half) :: [\<lambda>n xs. length xs < $2^{63}\text{\scriptsize{-}}1$] \shiftleft$\Asize$ \shiftleft\<rightarrow>\shiftleft $A_{\mathit{arl}}^d$ \shiftleft\<rightarrow>\shiftleft $A_{\mathit{arl}}$ %\label{exl:pref}%
\end{lstlisting}

\Cref{exl:append} defines \is{ls_app} (insert an element at the end of a list) and \is{half_nat} (divide a natural number in half).
Both are used in the definition of \is{app_half} in \Cref{exl:add}.
The standard library of \sepref has LLVM assertions for lists as array lists ($A_{\mathit{arl}}$), and natural numbers as 64-bit signed words ($\Asize$).

Now, \is{half_nat} can be refined to the LLVM program \is{rshift1}, which performs an efficient right bit shift.
\sepref represents refinements using parametric functor notation as in \Cref{exl:nref}.
The lemma states that \is{rshift1} and \is{half_nat} are related as follows:
If the inputs of \is{half_nat} (type \is{nat}) and \is{rshift1} (type \is{size}) are related via $\Asize$, then the outputs are related via $\Asize$.

\Cref{exl:aref} provides a refinement of the append operation following the same principle, only now with two inputs.
Moreover, the precondition \is{length xs < $2^{63} - 1$} limits the length of the input list \is{xs}.
The refinement relation only holds for sufficiently small lists.
Finally, the superscript $^d$ indicates that the input is destructively updated.
With these two rules registered to \sepref, we can automatically refine \is{app_half}.
We provide a signature to instruct \sepref to use the desired data structures: 
\begin{center}
\is{[\<lambda>n xs. length xs < $2^{63} - 1$] \shiftleft$\Asize$ \shiftleft\<rightarrow>\shiftleft $A_{\mathit{arl}}^d$ \shiftleft\<rightarrow>\shiftleft $A_{\mathit{arl}}$}.
\end{center}
\sepref can now automatically synthesize the LLVM program \is{arl_app_half} based on \is{rshift1} and \is{arl_app}.
Moreover, it proves the refinement relation in \Cref{exl:pref}.
\end{example}

\subsection{Floating-Point Extension of the Isabelle Refinement Framework}
\label{sec:fp_arith}

We extend the IRF with data refinements to floats.
While we can achieve this with error-bound estimation, the existing frameworks are cumbersome to use.
Instead, we propose to refine real numbers to lower bounding (\lbf) or upper bounding (\ubf) floats by employing safe rounding techniques~\cite{Hartmanns22}.
This allows us to perform a correctness proof on an algorithm that uses real numbers, and in a subsequent refinement replace them with \lbf/\ubf floats, implemented through \avx rounding modes.
We eliminate the need to compute error-bounds a-priori while still retaining the information relevant to proving the desired correctness properties.

Since the \ubf case is mostly symmetric to the \lbf case, we focus on \lbf floats.
We aim to construct a refinement relation that never produces \nan for the operations we support.
The main motivation here is that \nan is incomparable, rendering it incompatible with a framework that reasons about bounds.
Furthermore, the operations we support must preserve upper or lower bounds: the float $-2_f$ (the subscript $f$ denotes floats) is a lower bound of $1$, yet $-2_f \times -2_f = 4_f$ is not a lower bound of $1 \times 1 = 1$.
To ensure this property, we restrict ourselves to non-negative floats.

We define $\Rlb =$ \is{$\{$(fl,r). valof fl \<le> r \<and> \<not>is_nan fl \<and> valof fl >= 0$\}$}, the refinement relation that relates reals to \lbf floats, e.g. $(2_f, 3) \in \Rlb$, but $(-2_f, -1) \not\in \Rlb$, as $-2_f$ is negative.
Floats are \emph{pure}, \eg they do not need to allocate memory on the heap. This means that the assertion $\Alb$ is $\Rlb$ when provided an empty heap.

We now present a non-exhaustive set of operations supported by our framework, 
where we focus on the operations required for our use-case.

\paragraph{Fused Multiply-add.} The ternary operator \is{fma a b c = $a \times b + c$}~(fused multiply-add) yields a smaller floating-point rounding error compared to performing the two operations separately.
We name the \avx operation for \is{fma} that rounds towards negative infinity \is{fma_avx_lb} and prove the following refinement:
\begin{isalemma}[label=lst:fma]
lemma "(fma_avx_lb, fma) :: $\Alb$ \<rightarrow> $\Alb^{>0}$ \<rightarrow> $\Alb$ \<rightarrow> $\Alb$" %\hfill(\Cref{lst:fma})%
\end{isalemma}
This lemma states that if the inputs are $\lbf$, not \nan and non-negative, the output is also $\lbf$, not \nan and non-negative. 
Note that the result of $0_f * \infty_f$ is \nan.
We resolve this with the stricter assertion $\Alb^{>0}$ that only allows positive finite floats for the second argument.
In the case of II, the second argument is the transition probability of the MDP model, which is always positive.

\paragraph{Comparison.} Comparing two \lbf floats does not provide any information about comparing their real counterparts.
We can preserve information partially by implementing them as mixed operations, \eg, comparing \lbf to \ubf.
\begin{isalemma}[label=lst:leq]
definition "leq_sound a b = spec (\<lambda>r. r \<longrightarrow> a \<le> b)" %\hfill(\Cref{lst:leq})%
lemma "(leq_double, leq_sound) :: $\Aub$ \<rightarrow> $\Alb$ \<rightarrow> $A_{\mathit{bool}}$
\end{isalemma}
We use \is{spec} to introduce non-determinism as follows: the operation must return \is{False} if the comparison returns \is{False}, but can return anything otherwise.
For illustration of the latter case, consider the following 2 cases when comparing $2 \leq 5$. Case 1: $4_f$ is a \ubf of $2$, and $3_f$ is an \lbf of $5$, we have $4_f \not\leq 3_f$;
Case 2: $3_f$ is a \ubf of $2$, and $4_f$ is an \lbf of $5$, we have $3_f \leq 4_f$.
So both outcomes are possible after the refinement.

We use the comparison operation to implement the stopping criterion of II, in which case this partial information is sufficient.
For similar reasons, we also implement subtraction as a mixed operator (omitted here).

\paragraph{Min and Max.} It is possible to refine the minimum (\is{min}) and maximum (\is{max}) operations directly using comparisons.
We define the following refinement:
\begin{isalemma}[label=lst:min]
definition "min_double fl1 fl2 = if fl1 <= fl2 then fl1 else fl2" %\hfill(\Cref{lst:min})%
lemma "(min_double, min) :: $\Alb$ \<rightarrow> $\Alb$ \<rightarrow> $\Alb$""
\end{isalemma}

This refinement holds despite the fact that a comparison does not reveal anything about the bounding float number.
Consider the following case: $4_f$ is a lower bound of $5$ and $3_f$ is a lower bound of $6$.
Also, \is{min_double 4$_f$ 3$_f$ = 3$_f$} is a lower bound of \is{min 5 6 = 5}, even though the floating-point implementation returns the first argument, while the definition on reals returns the second argument.
The refinement works analogously for all combinations of \lbf/\ubf and \is{min}/\is{max}.

\paragraph{Constants.} We provide trivial refinements for the real number constants $0$ and $1$, which can be exactly represented as floating-point numbers.

\subsection{Refinement of Interval Iteration}
\label{sec:ii_refinement}
Using our floating-point extension to the IRF,
we derive an implementation of II using floats and directed rounding from the abstract specification in \Cref{sec:interval_iteration}.
The IRF allows us to reuse the correctness proofs of the abstract specification, and reason about the correctness of the implementation in isolation.
Through this separation of concerns we avoid directly proving the floating-point implementation correct.

\begin{wrapfigure}{r}{0.4\textwidth}
\vspace{-20pt}
\centering
\begin{tikzpicture}[scale=0.28]
\tikzstyle{every node}=[font=\normalsize]

\draw [ color=red, line width=2pt, dashed] (1.25,2) -- (13.75,2);
\node at (3.75,-1.25) [circle,fill,inner sep=1.5pt, color=olive] {};
\node at (6.25,-0.25) [circle,fill,inner sep=1.5pt, color=olive] {};
\node at (8.75,0.5) [circle,fill,inner sep=1.5pt, color=olive] {};
\node at (11.25,1) [circle,fill,inner sep=1.5pt, color=olive] {};
\node at (13.75,1.25) [circle,fill,inner sep=1.5pt, color=olive] {};
\draw [ color=olive, line width=2pt] (1.25,-3) -- (3.75,-1.25);
\draw [ color=olive, line width=2pt] (3.75,-1.25) -- (6.25,-0.25);
\draw [ color=olive, line width=2pt] (6.25,-0.25) -- (8.75,0.5);
\draw [ color=olive, line width=2pt] (8.75,0.5) -- (11.25,1);
\draw [ color=olive, line width=2pt] (11.25,1) -- (13.75,1.25);
\node at (3.75,7) [circle,fill,inner sep=1.5pt, color=olive] {};
\node at (6.25,5) [circle,fill,inner sep=1.5pt, color=olive] {};
\node at (8.75,3.75) [circle,fill,inner sep=1.5pt, color=olive] {};
\node at (11.25,3.25) [circle,fill,inner sep=1.5pt, color=olive] {};
\node at (13.75,3) [circle,fill,inner sep=1.5pt, color=olive] {};
\draw [ color=olive, line width=2pt] (1.25,9.5) -- (3.75,7);
\draw [ color=olive, line width=2pt] (3.75,7) -- (6.25,5);
\draw [ color=olive, line width=2pt] (6.25,5) -- (8.75,3.75);
\draw [ color=olive, line width=2pt] (8.75,3.75) -- (11.25,3.25);
\draw [ color=olive, line width=2pt] (11.25,3.25) -- (13.75,3);
\node at (3.75,-1.5) [circle,fill,inner sep=1.5pt, color=violet] {};
\node at (6.25,-0.75) [circle,fill,inner sep=1.5pt, color=violet] {};
\node at (8.75,-0.25) [circle,fill,inner sep=1.5pt, color=violet] {};
\node at (11.25,0) [circle,fill,inner sep=1.5pt, color=violet] {};
\node at (13.75,0) [circle,fill,inner sep=1.5pt, color=violet] {};
\node at (3.75,7.25) [circle,fill,inner sep=1.5pt, color=violet] {};
\node at (6.25,5.5) [circle,fill,inner sep=1.5pt, color=violet] {};
\node at (8.75,4.5) [circle,fill,inner sep=1.5pt, color=violet] {};
\node at (11.25,4) [circle,fill,inner sep=1.5pt, color=violet] {};
\node at (13.75,3.75) [circle,fill,inner sep=1.5pt, color=violet] {};
\draw [ color=violet, line width=2pt] (1.25,9.5) -- (3.75,7.25);
\draw [ color=violet, line width=2pt] (3.75,7.25) -- (6.25,5.5);
\draw [ color=violet, line width=2pt] (6.25,5.5) -- (8.75,4.5);
\draw [ color=violet, line width=2pt] (8.75,4.5) -- (11.25,4);
\draw [ color=violet, line width=2pt] (11.25,4) -- (13.75,3.75);
\draw [ color=violet, line width=2pt] (1.25,-3) -- (3.75,-1.5);
\draw [ color=violet, line width=2pt] (3.75,-1.5) -- (6.25,-0.75);
\draw [ color=violet, line width=2pt] (6.25,-0.75) -- (8.75,-0.25);
\draw [ color=violet, line width=2pt] (8.75,-0.25) -- (11.25,0);
\draw [ color=violet, line width=2pt] (11.25,0) -- (13.75,0);
\draw [line width=2pt] (1.25,-3) -- (1.25,9.5);
\draw [line width=2pt] (1.25,-3) -- (13.75,-3);
\node [font=\normalsize] at (7,0.9) {$lb$};
\node [font=\normalsize] at (7,3.8) {$ub$};
\node [font=\normalsize, rotate around={90:(0,0)}] at (0.5,3) {Value};
\node [font=\normalsize] at (7.5,-3.7) {Iterations};
\node [font=\normalsize] at (7,-1.7) {$\mathit{lb_f}$};
\node [font=\normalsize] at (7,6.5) {$\mathit{ub_f}$};
\node [font=\normalsize] at (3.5,2.7) {\is{\n}};
\end{tikzpicture}
\caption{The valuation of an MDP state over successive iterations: reals (yellow) vs. floats with safe rounding (purple).}
\label{fig:iiplot}
\vspace{-20pt}
\end{wrapfigure}

The plot in \Cref{fig:iiplot} shows a fictive run of II on both reals and their refinement to bounding floats.
In the long run, II on reals converges to the dashed red line \is{\n}.
The yellow lines denote the valuations of an MDP state using reals, \is{lb} starting from \is{lb_0} and \is{ub} from \is{ub_0}.

Implementing \is{lb} with lower-bounding floats (\Alb) yields the purple line $\mathit{lb_f}$ (similarly for \is{ub} using upper-bounding floats).
Note that the deviations are exaggerated on purpose in this example.

Formally, the following specification states soundness of II, \ie the outputs are lower and upper bounds of the actual reachability probabilities:
\begin{isadefinition}[label=lst:ii_spec]
definition "ii_inf_spec M = %\hfill(\Cref{lst:ii_spec})%
    spec (\<lambda>(x, y). \<forall>s \<in> states M. x s \<le> \n M s \<and> \n M s \<le> y s)
\end{isadefinition}

Convergence follows from \Cref{lst:bounds_diff}, but we do not include this in our correctness proof.
With our framework, this statement would be void: The float implementations of \is{x} and \is{y} are further apart than \is{x} and \is{y}, which are less than $\varepsilon$ apart.
As a first step towards the refinement to LLVM, we define II in the nres-monad (the \is{sup} case is analogous):
\begin{isadefinition}[numbers=left, label=lst:ii_def]
    definition "ii_gs_inf M L = %\label{line:def}% %\hfill(\Cref{lst:ii_def})%
        x \<leftarrow> lb$_0$ M; y \<leftarrow> ub$_0$ M; i \<leftarrow> 0; flag \<leftarrow> True;%\label{line:init}%
        while (i++ < L \<and> flag) ( %\label{line:while}%
            (x,y) \<leftarrow> F$_{\mathit{inf}}^{\,\mathit{gs}}$ M x y %\label{line:upd}%
            flag \<leftarrow> spec(\<lambda>x. True)) %\label{line:flag}%
        return (x,y) %\label{line:res}%"
\end{isadefinition}
We define \is{ii_gs_inf} in \Cref{line:def}. 
It takes as inputs an MDP \is{M}, and a maximal iteration count \is{L} to guarantee termination.
\Cref{line:init} initializes variables, most importantly the lower bounds \is{lb} and upper bounds \is{ub}.
The \is{flag} non-deterministically decides whether to abort the loop.
This is sound, because \is{ii_inf_spec} is satisfied after any number of iterations.
In each iteration, we first update the valuations according to a Gauss-Seidel variant of \is{F_inf} (\Cref{line:upd}):
We update \is{lb} and \is{ub} in-place, thereby we use the updated values already in the current iteration and converge faster.

The algorithm is now in a format ready for refinement proofs to LLVM.
Using the setup from \Cref{sec:interval_iteration} and \Cref{lst:bounds_preserved}, it is straightforward to prove that the algorithm refines the specification:
\begin{isatheorem}[label=lst:ii_refinement]
theorem "ii_gs_inf M L $\refcomp[]$ ii_inf_spec M" %\hfill(\Cref{lst:ii_refinement})%
\end{isatheorem}

\subsection{Refinement of the \mcsta Data Structure}
\label{sec:data_refinement}
The motivation behind refining II to LLVM code is to embed it into the model checker \mcsta from the \toolset~\cite{HartmannsH14}.
\mcsta is an explicit-state probabilistic model checker that also supports quantitative model checking of MDPs.
To avoid costly conversions of the MDP representation at runtime, we refine the \is{mdpc} data structure to the MDP data structure of \mcsta. This is a two-step process: First, we do a data refinement from \is{mdpc} to the sparse-matrix representation used by \mcsta based on HOL lists.
Then, we use \sepref to refine this data structure to LLVM. The sparse-matrix representation that we use is a 6-tuple:
\begin{center}\is{(St::nat list, Tr::nat list, Br::nat list, Pr::real list, u$_a$::nat, u$_t$::nat)}.\end{center}
For each state, \is{St} contains an index to \is{Tr}, pointing to the first transition (action) of the state.
Similarly, for each transition, \is{Tr} contains its first index \is{Br} and \is{Pr}, pointing to the first branch of the transition and its probability.
Finally, \is{Br} contains the target of the branch, pointing to \is{St}.
Additionally, \is{u$_a$} and \is{u$_t$} are the avoid and target states respectively.
\Cref{ex:mdpmcsta} illustrates this data structure.

\begin{example}
    \label{ex:mdpmcsta}
A possible representation of the MDP from \Cref{fig:mdp} is \is{St = [0,2,3,4,5]}, \is{Tr = [0,1,3,4,6,7]}, \is{Br = [2,0,1,0,1,3,3]}, \is{Pr = [1.0,0.9,0.1,1,0.5,0.5,1.0]}.
The index of the avoid and target state are stored in \is{u$_a$} and \is{u$_t$}. State $0$ has two actions: $\alpha$ and $\beta$ (transitions $0$ and $1$). Transition $1$ (action $\beta$) has two branches, $1$ and $2$, that lead to state $0$ and $1$ with probability $0.9$ and $0.1$ respectively.
\end{example}

\paragraph{Refinement Relation.}
We relate the abstract MDP type \is{mdpc} to the concrete data structure of \mcsta with the refinement relation \is{R$_M$} (definition omitted).
For example, \is{R$_M$} contains a tuple of the MDP of \Cref{fig:mdp} and \Cref{ex:mdpmcsta} as well as with each other instance of the data structure in \Cref{sec:data_refinement} along with the MDP it represents.
These lists present in the \isabelle model of the data structure can be directly refined to arrays of 64-bit integers (signed for compatibility with \mcsta).
Through composition we obtain the assertion $A_M$ from abstract MDP to LLVM.

\paragraph{Refinement of Operations.}
We use \sepref to refine the functions \is{lb_0}, \is{ub_0}, \is{F$_{\mathit{inf}}^{\,\mathit{gs}}$} and \is{flag} that are used by II.
Refining \is{lb_0} and \is{ub_0} to the concrete data structure is straightforward: we initialize an array and set entries to constants $0_f$ or $1_f$.
The floating-point refinement of \is{F$_{\mathit{inf}}^{\,\mathit{gs}}$} builds on \is{fma} and \is{min} (\is{max} for \is{F$_{\mathit{sup}}^{\,\mathit{gs}}$}) from \Cref{sec:fp_arith}.
Finally, we implement \is{flag} as follows: 
We compare the upper and lower bound, and set the flag if the difference is less than $\varepsilon$, specified by the user.
Using the above refinements for all operations in \is{ii_gs_inf},
we use \sepref to obtain an LLVM algorithm \is{ii_gs_inf_llvm} within \isabelle.

\subsection{Correctness Statement} 
\label{sec:hoare}

For the final step in our proof, we combine all of our proofs into one theorem.
We use a Hoare triple format, as it allows us to present a concise correctness statement of a program that is not cluttered by intermediary tools such as the parametric functor notation.
We show that our LLVM program \is{ii_gs_inf_llvm} satisfies the specification \is{ii_inf_spec}~(\Cref{lst:ii_spec}, unfolded in \Cref{lst:hoare}).
The resulting triple transitively combines \Cref{lst:ii_refinement} with the refinements from \Cref{sec:data_refinement}:
\begin{isatheorem}[numbers=left,label=lst:hoare]
    theorem "llvm_htriple %\hfill(\Cref{lst:hoare})%
      ($\Asize$ n n\_i ** $\Asize$ L L\_i ** $\Alb$ \<epsilon> \<epsilon>\_f ** $A_M$ M M\_i %\label{line:preassn}%
         ** \<up>(MDP_Reach M \<and> n+1 < max_size \<and> n = card (states M))) %\label{line:precond}%%\vspace{0.4em}%
      (ii_gs_inf_llvm L\_i n\_i \<epsilon>\_f M\_i) %\label{line:llvm}%%\vspace{0.4em}%
      (\<lambda>(lb\_f,ub\_f). \<exists>lb ub. A$_{\mathit{lb}}^{\mathit{out}}$ lb lb\_f ** A$_{\mathit{ub}}^{\mathit{out}}$ ub ub\_f %\label{line:pres}%
         ** \<up>(\<forall>s \<in> states M. lb s \<le> \n M s \<and> \n M s \<le> ub s)). %\label{line:post}%"
\end{isatheorem}

\Cref{line:preassn,line:precond} specify the preconditions, where
\Cref{line:preassn} concerns the input data: \is{n} and \is{L} are natural numbers implemented as 64-bit words \is{n\_i} and \is{L\_i}; \is{\<epsilon>} is a real number implemented as the float \is{\<epsilon>\_f} and \is{M} is the MDP implemented as \is{M\_i}. 
The separation conjunction \is{**} specifies that these implementations do not overlap on the heap.
\Cref {line:precond} is a boolean predicate lifted to separation logic using \is{\<up>}.
It states that \is{M} satisfies locale \is{MDP_Reach} (\Cref{lst:reduced_target}) and limits the number of states to the largest 64-bit number.

If these preconditions hold, a run of the algorithm (\Cref{line:llvm}) yields the arrays \is{lb$_f$} and \is{ub$_f$} that satisfy the postconditions in \Cref{line:pres,line:post}. 
These state that \is{lb$_f$} is a lower-bound implementation of \is{lb}, which is in turn a lower bound of \is{\n} (similarly for \is{ub}).
Note that the Hoare triple does not guarantee convergence of \is{lb$_f$} and \is{ub$_f$} to a single fixed point.
However, our experiments show that our implementation converges up to the limits of floating-point precision in practice.

\section{Experimental Evaluation}
\label{sec:implementation}
The LLVM code generator of the IRF exports the LLVM program \is{ii_gs_inf_llvm} for use in the LLVM compiler pipeline.
Additionally, it generates a C header that facilitates embedding the program into other software, such as \mcsta.

\paragraph{Integration with \mcsta.}
We integrate our verified implementation with \mcsta, replacing the existing unverified interval iteration algorithm.
This requires a small amount of unverified glue code to convert the MDP's probabilities into a floating-point representation:
\mcsta stores the probabilities in the MDP as 128-bit rationals (encoded as a pair of 64-bit integers representing the numerator and denominator).
Our MDP data structure \is{M\_i} expects two floats per branch, representing lower and upper bounds of the rational probability.
We convert the probabilities to 64-bit doubles by first converting the numerator and denominator to doubles and then performing two separate division operations, once rounding up and and once down.
We assume that the input MDP as produced by the \mcsta pipeline is well-formed.
If there were a bug in the parser that produces MDPs that are not well-formed according to the precondition of the Hoare triple, we would lose the formal guarantees.
As long as parts of the toolchain remain unverified, we rely on the correctness of the \mcsta implementation for the preprocessing.

\paragraph{Evaluation Questions.} 
We have formally verified in \isabelle that II with precise arithmetic computes lower and upper bounds (\is{lb} and \is{ub}), and eventually converges to the reachability probability.
However, two important questions remain:
First, verified implementations tend to be orders of magnitude slower than unverified ones~\cite{valIterIsabelle}.
Since we verified an implementation with efficient numerics down to LLVM, we expect much faster runtimes.
So how does our verified implementation compare to state-of-the-art tools in terms of performance?
Second, the floating-point outputs (\is{lb\_f} and \is{ub\_f}) provably provide conservative bounds.
Can we experimentally confirm that the algorithm converges in practice?

\paragraph{Setup.}
For the first question, we compare the runtime of our verified II implementation to its two unverified counterparts in \mcsta:
a \Csharp implementation with standard rounding (\textit{Modest} implementation) and a C implementation with safe rounding~\cite{Hartmanns22} (\textit{Safe} implementation).
The latter is similar to our verified LLVM implementation, also using \avx instructions for safe rounding.

We set the maximal iteration count to a high value ($10^7$) to ensure the computation is never terminated prematurely before convergence.
While we anticipate all benchmarks to converge within fewer iterations, this upper limit provides a safeguard.
We set the convergence threshold to $\varepsilon = 10^{-6}$.
Once the lower and upper bound differ by less than $\varepsilon$, the \textit{Verified}, \textit{Safe}, and \textit{Modest} implementations terminate.
We use all DTMC, MDP and PTA models of the Quantitative Verification Benchmark Set (QVBS)~\cite{BuddeHKKPQTZ20} that contain $10^6$ to $10^8$ states and need at least two iterations to converge to $\varepsilon$.
For our benchmarks, we consider both minimal and maximal reachability probabilities.
In total, this yields a benchmark set of 49 benchmark instances.
We execute all benchmarks on an Intel i9-11900K (3.5-5.3\,GHz) system with 128\,GB of RAM running Ubuntu Linux 22.04.

\begin{figure}[t]
  \centering
	\pgfplotstableread[col sep=comma]{
category,instance,x,y,
dtmc,brp.2048-64.p1,2.033333333333333,1.5,
dtmc,brp.2048-64.p2,2,1.5666666666666667,
dtmc,crowds.6-15.positive,0.6,0.6,
dtmc,crowds.5-20.positive,0.4,0.4,
dtmc,crowds.6-20.positive,2.3,2.6333333333333337,
dtmc,haddad-monmege.20-0.7.target,1.5,1.2,
mdp,beb.4-8-7.LineSeized,5.1,5.4,
mdp,beb.4-8-7.GaveUp,2.9333333333333336,3.4,
mdp,consensus.6-2.c2,6.3,6.2,
mdp,consensus.6-2.disagree,21.6,21,
mdp,csma.3-4.all_before_max,0.7,0.6,
mdp,csma.3-6.all_before_max,64,64,
mdp,csma.3-4.all_before_min,0.7,0.7,
mdp,csma.3-6.all_before_min,64,64,
mdp,csma.3-4.some_before,0.3,0.4,
mdp,csma.3-6.some_before,38,38.66666666666667,
mdp,echoring.50.MinFailed,0.3,0.3,
mdp,echoring.100.MinFailed,1.1,1,
mdp,echoring.50.MinOffline1,0.3,0.25,
mdp,echoring.100.MinOffline1,1.1,0.9,
mdp,echoring.50.MaxOffline1,0.3,0.25,
mdp,echoring.100.MaxOffline1,1.1,0.9,
mdp,echoring.50.MinOffline2,0.3,0.25,
mdp,echoring.100.MinOffline2,1,0.9,
mdp,echoring.50.MaxOffline2,0.3,0.25,
mdp,echoring.100.MaxOffline2,1,0.9,
mdp,echoring.50.MinOffline3,0.3,0.25,
mdp,echoring.100.MinOffline3,1,0.9,
mdp,echoring.50.MaxOffline3,0.3,0.25,
mdp,echoring.100.MaxOffline3,1.0333333333333334,0.9,
mdp,firewire.true-3-600.deadline,0.25,0.25,
mdp,firewire.true-3-800.deadline,0.3,0.3,
mdp,firewire.true-36-600.deadline,10.3,11.7,
mdp,ij.20.stable,27.3,28.9,
mdp,wlan_dl.2-80.deadline,0.25,0.25,
mdp,wlan_dl.3-80.deadline,0.25,0.25,
mdp,wlan_dl.4-80.deadline,0.5,0.5,
mdp,wlan_dl.5-80.deadline,0.7,0.7,
mdp,wlan_dl.6-80.deadline,0.7,0.7,
mdp,zeroconf.20-8-False.correct_max,0.4,0.5,
mdp,zeroconf.1000-8-False.correct_max,0.5,0.5,
mdp,zeroconf.20-8-False.correct_min,0.5666666666666667,0.6,
mdp,zeroconf.1000-8-False.correct_min,0.6,0.6,
pta,brp-pta.64-12-32-256.P_1,24.433333333333334,23.733333333333334,
pta,brp-pta.64-12-32-256.P_2,23,22.4,
pta,brp-pta.64-12-32-256.P_3,23.7,22.966666666666665,
pta,firewire-pta.30-7500.eventually,18.966666666666665,16.833333333333332,
pta,wlan-large.2.P_min,0.4,0.4,
pta,wlan-large.2.P_max,0.6,0.6,

}\dataiitimemcstalpModestIImcstalpVerifiedII

\pgfplotstableread[col sep=comma]{
category,instance,x,y,
dtmc,brp.2048-64.p1,1.7,1.5,
dtmc,brp.2048-64.p2,1.7,1.5666666666666667,
dtmc,crowds.6-15.positive,0.6,0.6,
dtmc,crowds.5-20.positive,0.43333333333333335,0.4,
dtmc,crowds.6-20.positive,2.9,2.6333333333333337,
dtmc,haddad-monmege.20-0.7.target,1.1,1.2,
mdp,beb.4-8-7.LineSeized,5.4,5.4,
mdp,beb.4-8-7.GaveUp,3.4,3.4,
mdp,consensus.6-2.c2,5.5,6.2,
mdp,consensus.6-2.disagree,18.5,21,
mdp,csma.3-4.all_before_max,0.6,0.6,
mdp,csma.3-6.all_before_max,64,64,
mdp,csma.3-4.all_before_min,0.7,0.7,
mdp,csma.3-6.all_before_min,64,64,
mdp,csma.3-4.some_before,0.4,0.4,
mdp,csma.3-6.some_before,39.03333333333333,38.66666666666667,
mdp,echoring.50.MinFailed,0.25,0.3,
mdp,echoring.100.MinFailed,0.9,1,
mdp,echoring.50.MinOffline1,0.25,0.25,
mdp,echoring.100.MinOffline1,0.8,0.9,
mdp,echoring.50.MaxOffline1,0.25,0.25,
mdp,echoring.100.MaxOffline1,0.8,0.9,
mdp,echoring.50.MinOffline2,0.25,0.25,
mdp,echoring.100.MinOffline2,0.8,0.9,
mdp,echoring.50.MaxOffline2,0.25,0.25,
mdp,echoring.100.MaxOffline2,0.8,0.9,
mdp,echoring.50.MinOffline3,0.25,0.25,
mdp,echoring.100.MinOffline3,0.8,0.9,
mdp,echoring.50.MaxOffline3,0.25,0.25,
mdp,echoring.100.MaxOffline3,0.8,0.9,
mdp,firewire.true-3-600.deadline,0.25,0.25,
mdp,firewire.true-3-800.deadline,0.3,0.3,
mdp,firewire.true-36-600.deadline,11.133333333333333,11.7,
mdp,ij.20.stable,28.733333333333334,28.9,
mdp,wlan_dl.2-80.deadline,0.25,0.25,
mdp,wlan_dl.3-80.deadline,0.25,0.25,
mdp,wlan_dl.4-80.deadline,0.5,0.5,
mdp,wlan_dl.5-80.deadline,0.7,0.7,
mdp,wlan_dl.6-80.deadline,0.7,0.7,
mdp,zeroconf.20-8-False.correct_max,0.5,0.5,
mdp,zeroconf.1000-8-False.correct_max,0.5,0.5,
mdp,zeroconf.20-8-False.correct_min,0.6,0.6,
mdp,zeroconf.1000-8-False.correct_min,0.6,0.6,
pta,brp-pta.64-12-32-256.P_1,21.466666666666665,23.733333333333334,
pta,brp-pta.64-12-32-256.P_2,20.133333333333333,22.4,
pta,brp-pta.64-12-32-256.P_3,20.733333333333334,22.966666666666665,
pta,firewire-pta.30-7500.eventually,15.366666666666667,16.833333333333332,
pta,wlan-large.2.P_min,0.4,0.4,
pta,wlan-large.2.P_max,0.6,0.6,

}\dataiitimemcstalpSafeIImcstalpVerifiedII

%
%
\begin{tikzpicture}
  \begin{axis}[
    width=\scatterplotsize,
    height=\scatterplotsize,
    axis equal image,
    xmin={0.25}, ymin={0.25}, xmax={85.12}, ymax={85.12},
    xmode=log, ymode=log,
    axis x line=bottom,
    axis y line=left,
    xtick={0.25,0.5,1,2,4,8,16,32},
    xticklabels={\hspace{-5pt}${\leq}\,0.25$,$\mathstrut 0.5$,$\mathstrut 1$,$\mathstrut 2$,$\mathstrut 4$,$\mathstrut 8$,$\mathstrut 16$,$\mathstrut 32$},
    ytick={0.25,0.5,1,2,4,8,16,32},
    yticklabels={${\leq}\,0.25$,$\mathstrut 0.5$,$\mathstrut 1$,$\mathstrut 2$,$\mathstrut 4$,$\mathstrut 8$,$\mathstrut 16$,$\mathstrut 32$},
    xlabel={\strut{}Modest time (s)},
    xlabel style={font=\scriptsize,yshift=5pt},
    ylabel={\strut{}Verified time (s)},
    ylabel style={font=\scriptsize,yshift=-12pt},
    yticklabel style={font=\scriptsize},
    xticklabel style={font=\scriptsize},
    legend pos=north east,
    legend columns=-1,
    legend style={nodes={scale=0.75, transform shape},anchor=north east,inner sep=1.5pt,yshift=16pt,xshift=3.5pt},
    legend cell align={left},
    extra x ticks = {64},
    extra x tick labels = {$\mathstrut{\geq}\,64$},
    extra x tick style = {grid = major},
    extra y ticks = {64},
    extra y tick labels = {$\mathstrut{\geq}\,64$},
    extra y tick style = {grid = major}
  ]
  \addplot[
    scatter,
    only marks,
    scatter/classes={dtmc={mark=square*,color1,mark size=1.25}, mdp={mark=diamond*,color2,mark size=1.75}, pta={mark=triangle*,color3,mark size=1.75}},
    scatter src=explicit symbolic
  ]%
  table[x={x}, y={y}, meta={category}]{\dataiitimemcstalpModestIImcstalpVerifiedII};
  \legend{DTMC, MDP, PTA}
  \addplot[no marks] coordinates {(0.25,0.25) (64,64)};
  \addplot[no marks, densely dotted] coordinates {(0.5*0.25,0.25) (0.5*64,64)};
  \addplot[no marks, densely dotted] coordinates {(0.25,0.5*0.25) (64,0.5*64)};
  \end{axis}
\end{tikzpicture}%
%
%
\begin{tikzpicture}
  \begin{axis}[
    width=\scatterplotsize,
    height=\scatterplotsize,
    axis equal image,
    xmin={0.25}, ymin={0.25}, xmax={85.12}, ymax={85.12},
    xmode=log, ymode=log,
    axis x line=bottom,
    axis y line=left,
    xtick={0.25,0.5,1,2,4,8,16,32},
    xticklabels={\hspace{-5pt}${\leq}\,0.25$,$\mathstrut 0.5$,$\mathstrut 1$,$\mathstrut 2$,$\mathstrut 4$,$\mathstrut 8$,$\mathstrut 16$,$\mathstrut 32$},
    ytick={0.25,0.5,1,2,4,8,16,32},
    yticklabels={${\leq}\,0.25$,$\mathstrut 0.5$,$\mathstrut 1$,$\mathstrut 2$,$\mathstrut 4$,$\mathstrut 8$,$\mathstrut 16$,$\mathstrut 32$},
    xlabel={\strut{}Safe time (s)},
    xlabel style={font=\scriptsize,yshift=5pt},
    ylabel={\strut{}Verified time (s)},
    ylabel style={font=\scriptsize,yshift=-12pt},
    yticklabel style={font=\scriptsize},
    xticklabel style={font=\scriptsize},
    legend pos=north east,
    legend columns=-1,
    legend style={nodes={scale=0.75, transform shape},anchor=north east,inner sep=1.5pt,yshift=16pt,xshift=3.5pt},
    legend cell align={left},
    extra x ticks = {64},
    extra x tick labels = {$\mathstrut{\geq}\,64$},
    extra x tick style = {grid = major},
    extra y ticks = {64},
    extra y tick labels = {$\mathstrut{\geq}\,64$},
    extra y tick style = {grid = major}
  ]
  \addplot[
    scatter,
    only marks,
    scatter/classes={dtmc={mark=square*,color1,mark size=1.25}, mdp={mark=diamond*,color2,mark size=1.75}, pta={mark=triangle*,color3,mark size=1.75}},
    scatter src=explicit symbolic
  ]%
  table[x={x}, y={y}, meta={category}]{\dataiitimemcstalpSafeIImcstalpVerifiedII};
  
  \addplot[no marks] coordinates {(0.25,0.25) (64,64)};
  \addplot[no marks, densely dotted] coordinates {(0.5*0.25,0.25) (0.5*64,64)};
  \addplot[no marks, densely dotted] coordinates {(0.25,0.5*0.25) (64,0.5*64)};
  \end{axis}
\end{tikzpicture}%
	\vspace{-10pt}
	\caption{Comparison of elapsed runtime for completing Interval Iteration}
	\label{fig:iitime}
\end{figure}

\paragraph{Results.}
\Cref{fig:iitime} compares the runtimes of the three implementations.
Each point $\tuple{x, y}$ in a plot indicates that, on one benchmark instance, the implementation on the horizontal axis took $x$ seconds while the \textit{Verified} implementation took $y$ seconds.
We note that none of our benchmark instances reached the timeout of 10 minutes.
We provide the runtimes for the II algorithm only.
Other steps are performed equally by \mcsta such as state space exploration or min/max-reduction.

We see that the \textit{Verified} implementation matches the unverified ones in terms of performance.
Thus, as far as this benchmark set can show, the answer to the first question is that we have achieved comparable performance to a state-of-the-art unverified tool with a fully verified implementation. 
We also observed that the \textit{Verified} implementation does not reach the maximal iteration count on any instance, \ie it always converged up to $\varepsilon$, indicating that the second question can also be answered affirmatively.

Additionally, we did not find any significant discrepancies in the raw data output\ifthenelse{\boolean{includeAppendix}}{(\Cref{sec:raw_data})}{}.
The number of iterations to convergence is equal for all instances except for the Haddad-Monmege model~\cite{HaddadM18} between \textit{Verified} and \textit{Modest}.
The exception is not surprising, as this model is designed to converge very slowly, increasing the influence of floating-point errors.
We also compared the computed results.
Due to using different rounding modes, the results of \textit{Verified} and \textit{Modest} show differences well within $\varepsilon$.
Despite the fact that \textit{Verified} and \textit{Safe} both implement safe rounding, their outputs still differ by minimal amounts on the order of $10^{-20}$.
This may be caused by floating-point operations being non-associative, so a different order of operations may yield different outputs.

In terms of memory consumption, \textit{Verified} and \textit{Safe} use almost the same amount of RAM, but use on average 20\% more RAM than \textit{Modest}.
Since the memory consumption is very similar for \textit{Verified} and \textit{Safe}, we suspect that these differences come from garbage collection effects caused by \textit{Verified} and \textit{Safe} being native code called from within a tool otherwise running in the managed \Csharp runtime, as opposed to the purely-\Csharp \textit{Modest} implementation.

\section{Discussion}
\label{sec:conclusion}

We provide a framework to verify floating-point LLVM programs which exploits the parametricity principle~\cite{theoremsFreeWadler} of the IRF by consistently applying directed rounding to floating-point values.
Our approach eliminates complex reasoning about error bounds to prove useful correctness properties of the implementation.
We are confident that our framework may be applied to other iterative algorithms that compute lower and upper bounds.
The latter is strictly necessary, as---with our approach---we do not a-priori know the size of the error.
However, if both bounds converge to a single value, the floating-point error is guaranteed to be smaller than the distance between the bounds.
Thus we can determine an a-posteriori error-bound.

We applied our extension to the IRF to the formal verification of the interval iteration algorithm in \isabelle---all the way to a correct implementation. 
Our developments formally prove that the algorithm computes lower and upper bounds for the reachability probabilities (soundness) and converges to a single fixed point (completeness) when using real numbers.
Furthermore, we show that soundness is preserved if we implement the algorithm using floating-point arithmetic with safe rounding.
Our proofs culminate in a single statement, presented as a Hoare triple, leaving no gaps in the link between the specification and the implementation in LLVM.

Finally, we extract verified LLVM code from our formalization and embed it in the \mcsta model checker of the Modest toolset.
We experimentally verify that our implementation converges in practice and is competitive with manually implemented, optimized, and unverified counterparts.
This is an important step towards a fully verified probabilistic model checking pipeline.

We also present our approach as an alternative to the bottom-up approach of building a verified model checker from scratch.
With our top-down approach, the full functionality of the model checker is available to the user, possibly in cross-usage with verified components.
Verified components are integrated with the model checker incrementally as drop-in replacements for unverified components, designed with competitive performance in mind.

In a next step, we plan to complete the verified II backend for \mcsta, the missing part being an efficient verified implementation of the transformations to reduced MDPs.
For this purpose, we aim to build on recent advancements that provide a verified and efficiently executable MEC decomposition algorithm in \isabelle~\cite{HartmannsKL24}.

\paragraph{Verification vs.\ Certification for II.}
An alternative to verifying II is to verify a \emph{certifier} that,
given the result from an unverified implementation of II plus a compact \emph{certificate}, can efficiently check that the result is indeed correct.
The advantage of certification is that the unverified implementation of II can be improved in an agile way independent of the certifier.
Also, since the certifier is (presumably) simpler than II, it should be easier to verify, including the verification of a high-performance implementation.
One possible certification scheme for II is using the Park induction check of the optimistic value iteration algorithm~\cite{HartmannsK20}, \ie performing one iteration of II and checking if the lower/upper bound does not decrease/increase for any state.
This could confirm that \emph{some} fixed point lies between the bounds computed by II.
This certification scheme would require as input
(1)~the final lower and upper bound for all states,
(2)~the full MDP, and
(3)~\emph{a certificate that the MDP is reduced}.
Without the latter, one can not be sure that the certified fixed point corresponds to the reachability probability, i.e.\ that it is the least fixed point.
One challenge with this scheme is that, unlike our verified implementation of II, it cannot guarantee that \emph{all} fixed points lie between the bounds.
Also, since the entire reduced MDP is input to the certificate checker, the certificate checking performance might be fundamentally limited.




\paragraph{Data Availability.} Proofs and benchmarks presented in this paper are archived and available at DOI
\href{https://doi.org/10.4121/bf0fef24-4f0f-4de6-a58d-07b9ba601804}{10.4121/bf0fef24-4f0f-4de6-a58d-07b9ba601804}~\cite{PaperArtifact}.

\paragraph{Disclosure of Interests.} The authors have no competing interests to declare that are relevant to the content of this article.

%
%
%
\bibliographystyle{splncs04}
\bibliography{paperA,paperB}
%

\end{document}